\documentclass[]{aa}

\newcommand{\msun}{\mbox{$\rm M_{\odot}$}}

\newcommand{\zsun}{\mbox{$\rm Z_{\odot}$}}

\usepackage{natbib}
\usepackage{graphics}

\begin{document}

\title{Periodic bursts of Star Formation in Irregular Galaxies}
\titlerunning{Bursts of SF in Irregulars}

\author{Pelupessy F.I. 
\inst{1} \and
        van der Werf P.P.
\inst{1} \and
	Icke V. 
\inst{1}
}
\authorrunning{Pelupessy, van der Werf, Icke}

\institute{ 
 {Leiden Observatory, Leiden University, 
  PO Box 9513, 2300 RA Leiden, The Netherlands} }
  
\date{ }

\abstract{ 

We present N-body/SPH simulations of the evolution of an isolated dwarf 
galaxy including a detailed model for the ISM, star formation and stellar 
feedback. Depending on the strength of the feedback, the modelled dwarf galaxy
shows periodic or quasi-periodic bursts of star formation of moderate 
strength. The period of the variations is related to the dynamical timescale,
of the order of $1.5~10^8$ yr. We show that the results of these simulations 
are in good agreement with recent detailed observations of dwarf irregulars (dIrr) and
that the peculiar kinematic and morphological properties of these objects,
as revealed by high resolution HI studies, are fully reproduced. We discuss these 
results in the context of recent surveys of dwarf galaxies and point out that if 
the star formation pattern of our model galaxy is typical for dwarf irregulars this
could explain the scatter of observed properties of dwarf galaxies. Specifically, 
we show that the time sampled distribution of the ratio between the instanteneous 
star formation rate (SFR) and the mean SFR is similar to that 
distribution in observed sample of dwarf galaxies.  
 
}  

\maketitle

\section{Introduction}

 The nature of the processes regulating star formation in irregular galaxies is 
 poorly understood. Whereas there is at least some understanding of 
 star formation in regular spiral galaxies, this is less so for irregulars. For 
 spiral galaxies the guiding observation that the star formation rate (SFR) is  
 related to the gas surface density by the Schmidt law has given rise to a number
 of competing theories that reproduce the general features of star formation in large spiral 
 galaxies (Elmegreen~\citeyear{E02}, Dopita \& Ryder~\citeyear{DR94}). These systems seem to be 
 regulated by large scale gravitational instabilities. Star formation in irregular
 galaxies has proven to be more difficult to understand. Irregulars have a widely varying 
 SFR, spanning 4 orders of magnitude for the normalized SFR/area (Hunter~\citeyear{H97}),
 possibly due to the fact that gas thermodynamics, governed by varying heating and cooling
 processes, plays the decisive role (Elmegreen~\citeyear{E02}). But why do some irregulars 
 have very high SFRs relative to their mass, while others hardly show any activity? Are 
 there any intrinsic properties of the galaxies that can explain this disparity between
 SFRs or do all dwarf galaxies exhibit episodes of high star formation? 
  
 In recent years a number of studies have highlighted these questions by
 investigating samples of dwarf galaxies and comparing their properties as derived 
 from photometry, HI and H$\alpha$ observations. Van Zee (\citeyear{Z00}, \citeyear{Z01})
 investigated a sample of isolated dwarf galaxies and found no strong correlation 
 between star formation and independent physical parameters. Hunter et al. (\citeyear{HEB98}) 
 tested different regulating processes, amongst which disk instabilities, thermal and shear 
 regulated star formation, but found that none could explain patterns of star formation. 
 Stil (\citeyear{S99}) investigated the relation between star formation and HI gas kinematics.  
 
 On the other hand, detailed studies of a number of nearby dwarf galaxies have fully revealed 
 the complex structure of the ISM in these systems. High resolution aperture synthesis 
 mapping (e.g. Kim et al.~\citeyear{KSSD98}, Wilcots \& Miller~\citeyear{WM98}, Puche et
 al.~\citeyear{PWBR92}, Walter \& Brinks~\citeyear{WB01}) of their HI has shown the  
 interstellar medium (ISM) of these dwarfs to be a frothy structure, with holes of varying 
 sizes, shells and filaments, even extending far beyond the optical radius. From velocity dispersion 
 studies (Young et  al.~\citeyear{YL97}) the presence of cold and warm neutral components predicted by the 
 two phase model for the ISM (Field~\citeyear{F65}) has been deduced. Comparison with 
 H$\alpha$ and UV observations shows that the dense walls of these holes are the sites 
 of star formation (Walter et al.~\citeyear{WT01}), and suggest 'chains' of 
 successive star forming sites (Stewart et al.~\citeyear{SFB00}). The cause of the holes in 
 the HI distribution seems to be the energy input from ionizing radiation, stellar winds 
 and supernovae, although Rhode et al.~(\citeyear{RSWR99}) and Efremov et al.~(\citeyear{EEH98}) 
 discuss other possible scenarios.     
  
 Together these two types of observations have painted a picture of the complex interaction 
 between star formation and the ISM of these systems that is formidable to capture 
 theoretically. Some early attempts have been made to understand star formation qualitatively 
 by the application of stochastic self propagating star formation (SSPSF) models to dwarf 
 galaxies (Gerola \& Seiden~\citeyear{GSS80}, Comins~\citeyear{C83}). While they probably 
 capture some general characteristics of star formation, they are phenomenological and do not 
 include the underlying physics of the ISM.
 
 Efforts to investigate the influence of star formation on the ISM of dwarf galaxies have 
 mainly concentrated on the effects of large (central) bursts and on questions concerning
 the ejection of gas and the distribution of metals(e.g. Mac-Low \& Ferrara~\citeyear{MF99}, 
 Mori et al.~\citeyear{MYTN97}). Recently there also  have been some simulations adressing 
 the question of survival of small galaxies (Mori et al.~\citeyear{MFM02}). Generally these
 simulations have not tried to set up a self consistent model for the ISM and star formation, 
 but prescribed a certain SFR. 
 
 The importance of a good model for the ISM and feedback has been recognized by a number of authors.
 Andersen \& Burkert~(\citeyear{AB00}) formulated an extensive model for the ISM in terms of a phenomenological
 model for the interstellar clouds. Their model showed self regulation of the SFR and they found moderate 
 fluctuations in SFR. Berczik \& Hensler~(\citeyear{BHTS03}) incorporated such a cloud model into a chemodynamical
 galaxy evolution code. Semelin and Combes~(\citeyear{SC02}) formulated a model with similar characteristics,
 representing clouds by sticky particles, but did not apply these to dwarf galaxies.
 Springel \& Hernquist~(\citeyear{SH03}) formulated a subgrid model for the multiphase interstellar medium, 
 producing a quiescent self-regulating ISM. However, relatively little effort has been directed towards resolving 
 the normal evolution of irregular dwarfs and providing the connection with detailed studies of single systems 
 and extensive unbiased samples. Nevertheless dwarf galaxies are good test systems for exploring star 
 formation in galaxies: they are dynamically simple systems in the sense that they do not exhibit spiral density 
 waves or shear. Furthermore their small size means that simulations can follow the various physical processes at 
 finer linear and density scales.\emph{ As the small scale physics of star 
 formation and feedback presumably do not differ between normal and dwarf galaxies, we can use results 
 obtained from these simulations and apply the same methods to larger systems}. Here we present 
 results of a simulation of the evolution of a normal dwarf irregular galaxy including a detailed
 model for the ISM, star formation and feedback. The distinguishing characteristics of this
 work are that the model for the ISM we employ does not explicitly postulate the presence of
 a two phase medium, rather it forms it as a result of the physics of the model. Furthermore 
 we take special care in formulating a star formation model that is solely based on the Jeans
 instability, and we formulate a feedback scheme that gives us unambiguous control over the 
 strength of the feedback. We will discuss the results of the simulation both in relation to
 detailed observations of comparable single systems, as well as in the context 
 of recent surveys of dwarf galaxies. 
 
\section{Method}

 We employ an N-body/SPH code for the evolution of a general astrophysical fluid on galactic
 scales, extended from TreeSPH (Hernquist \& Katz~\citeyear{HK89}), to simulate the evolution 
 of an isolated dwarf galaxy. We use the conservative SPH formulation of Springel \& Hernquist
 (\citeyear{SH02}). Main features of our code are: a realistic model for the ISM solving for 
 the ionization and thermal balance for the neutral and ionized components of the ISM, star 
 formation based on a gravitational instability model for clouds, and a new method of 
 including feedback for SPH. We will summarize the features of the code with an emphasis on 
 the aspects most relevant for the present work.  

\subsection{Model for the ISM}

\begin{table*}
\begin{tabular}{l c l}
\hline
 process &             comment &             ref.    \\
 \hline
 \emph{heating} & & \\
 ~~Cosmic Ray & ionization rate $\rm \zeta_{CR}=1.8~10^{-17}~s^{-1}$   & 1  \\
 ~~Photo Electric & FUV field from stars & 1  \\
\hline
 \emph{cooling} & & \\
 ~~$e$,H$_0$ impact   & H,He,C,N,O,Si,Ne,Fe & 2,4 \\
\hline
 \emph{ionization} & & \\
 \emph{~\& recombination} & & \\
 ~~UV & ionization assumed for & \\
      & species with $E_i< 13.6 \rm{eV}$ &  \\
 ~~Cosmic Ray & H, He only; primary & 1 \\
            & \& secondary ionizations & \\
 ~~Collisional & H, He only & 3 \\
 ~~Radiative recombination & H, He only & 3 \\	    
 ~~CIE & assumed for metals & \\ 
\hline	     	 	 
\end{tabular} 

\caption[]{
Overview of the processes included in the ISM model used. For H and He ionization equilibrium is
explicitly calculated, for other elements collisional ionization equilibrium (CIE) is assumed. 
Both the heating and cooling strongly depend on the ionization fraction$x_e$. Exact expressions adopted 
for the various processes can be found in: 
1) Wolfire et al.~\citeyear{WHMTB95}, 2) Raga et al.~\citeyear{RML97} , 
3) Verner \& Ferland~\citeyear{VF96}, 4) Silva \& Viegas~\citeyear{SV01} 
} 
\label{ism_tbl}

\end{table*} 

 Our model for the ISM is, although simplified, qualitatively similar to the model for the 
 Cold Neutral Medium (CNM) and Warm Neutral medium (WNM) of Wolfire et al.~(\citeyear{WHMTB95},~\citeyear{WMHT03}). 
 We consider a gas with arbitrary but fixed chemical abundances $X_i$, scaled to the target 
 metallicity from the solar abundances of Grevesse \& Sauval~(\citeyear{GS98}). 
 We solve for the ionization and thermal evolution of the gas.  The various processes included are
 given in Table~\ref{ism_tbl}. A similar model to that employed here
 was used by Gerritsen \& Icke~(\citeyear{GI97}) and Bottema~(\citeyear{B03}) for galaxy simulations. 
 The main differences are the following: we use more accurate cooling, that is calculated in 
 accordance with the chemical composition, we have included a solver for the ionization balance, and
 we use the full photoelectric heating efficiency as given in Wolfire et al.~(\citeyear{WHMTB95}).
 Gerritsen \& Icke~(\citeyear{GI97}) found that the structure of the resulting ISM depended strongly on the ionization 
 fraction they assumed, as this strongly influences the cooling. We do not have to assume an 
 ionization fraction, as we calculate it (on the other hand, we do assume a cosmic ray ionization rate 
 that is poorly constrained). Our use of the full heating efficiency means that FUV heating will become
 less efficient for high radiation fields, due to grain charging. In our model supernova (SN) 
 heating is more important in regulating star formation than it was for Gerritsen \& Icke~(\citeyear{GI97}).

\begin{figure}
\resizebox{\hsize}{!}{\includegraphics{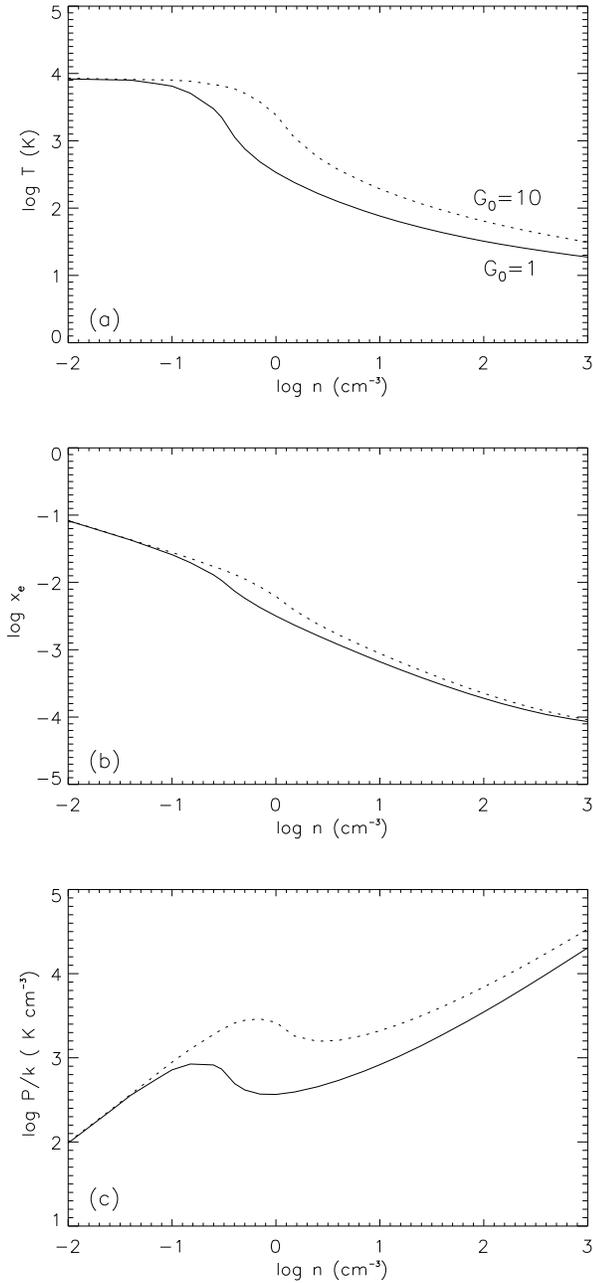}}
\caption{ Overview of the ISM model: Equilibrium plots of (a) temperature T,
(b) electron fraction x$_e$ and (c) pressure P as a function of density n for two different values of
the UV field G$_0$, drawn line: G$_0=1$, dotted: G$_0=10$ ( $\times 1.6~10^{-3}~{\rm ergs~cm^{-2}~s^{-1}}$ )}
\label{ismfig}
\end{figure}

 A concise overview of the ISM model is given in Fig.~\ref{ismfig}. The plots in this figure show that as
 density varies, the equilibrium state of the gas changes from a high temperature/high ionization state 
 ($\rm T=10k~K$, $\rm x_e \approx 0.1$) at low densities, to a low temperature/low ionization state 
 ($\rm T<100~K$, $\rm x_e<10^{-3}$) at high densities. In between is a density domain where the negative slope of
 the P-n relation indicates that the gas is unstable to isobaric pressure variations, the classic thermal 
 instability (Field~\citeyear{F65}). The shape of these curves and hence the exact densities of the thermal 
 instability vary locally throughout the simulation according to the conditions of UV and supernova heating. 
 The gas in the simulation may be out of equilibrium, although the timescales for reaching equilibrium are generally
 short, $\rm <10^6~yr$. In principle the cooling properties of the gas depend on the local chemical composition. 
 In practice, only small metallicity gradients are observed in dwarfs (Pagel \& Edmunds~\citeyear{PE81}), so we take 
 constant metallicity($\rm Z=0.2~\zsun$). Potentially more serious is the fact that we assume a constant cosmic ray 
 ionization rate $\zeta$ throughout the galaxy. The low energy ($ \approx 100~{\rm MeV}$) cosmic 
 rays that are important for heating and ionization of the ISM, have relatively short mean free paths( $< 10$ pc) so
 probably $\zeta$ will vary substantially across a galaxy. However, the exact sources, let alone production 
 rates, of these low energy cosmic rays are not well known. Hence a satisfying model for the distribution of
 cosmic rays is difficult to formulate. We take a 'standard' value of $\zeta=1.8~10^{-17}~{\rm s}^{-1}$.
 For this value cosmic ray heating only becomes important in the outer parts of the galaxy (it may 
 be substantially too low, see McCall et al.~\citeyear{MHS03}). 
         
 The FUV luminosities of the stellar particles, which are needed to calculate the local FUV 
 field used in the photoelectric heating, are derived from Bruzual \& Charlot (\citeyear{BC93},
 and updated) population synthesis models for a Salpeter IMF with 
 cutoffs at 0.1~\msun  and 100~\msun. In the present work we do not account for dust 
 extinction of UV light.
   
\subsection{Star formation and Feedback}

 We use the star formation recipe of Gerritsen \& Icke~(\citeyear{GI97}). A region is 
 considered unstable to star formation if the local Jeans mass $\rm M_J$ is smaller than 
 the mass of a typical molecular cloud $\rm M_{ref} \approx 10^5 \msun$. The rate of star 
 formation is set to scale with the local free fall time: 
 $\rm \tau_{sf}=f_{sf} t_{ff}= \frac{f_{sf}}{\sqrt{4 \pi G \rho}}$.  The delay factor 
 $\rm f_{sf}$ is uncertain, but from observations a value  $\rm f_{sf} \approx 10$ seems 
 reasonable (Zuckerman \& Palmer~\citeyear{ZE74}). Once a gas particle is determined to
 be forming stars, a fraction $\rm \epsilon_{sf}=0.25$ of the mass is converted to stars. This sets
 a minimum to the star formation efficiency (a neighbourhood of gas particles of about 64
 particles that has become capable of star formation will thus form at least a fraction $.25/64=.004$
 of stars). The actual efficiency of star formation is determined by the number of stars needed 
 to quench star formation locally by the UV and SN heating and is determined by the cooling
 properties of the gas and the energy input from the stars.
 
 The recipe is certainly not unique, we have based our choice on the following considerations:
 it is simple and based on presence of substructure and the driving role of gas self gravity, 
 and it reproduces the Schmidt law without actually imposing it( Gerritsen \& Icke~\citeyear{GI97}).
 An additional advantage is the fact that the Jeans mass criterion for star formation also 
 prevents the simulation from violating the resolution requirements for self gravitating 
 SPH (Bate \& Burkert~\citeyear{BB97}, Whitworth~\citeyear{W98} ). It does assume substructure 
 to be present (actually its assumption is more restrictive still, namely that substructure is 
 mainly at $\rm M_{ref}$ sized clouds, but this is not overly important). For this type of 
 simulation one is always restricted by the limited ability to follow the starforming 
 process, so we are forced to adopt a phenomenological description at some level.
 Another concern that could be adressed is that the SF according to this recipe is independent
 of metallicity, other than that induced by the metallicity dependence of the cooling.
 Furthermore, we assume the stellar IMF to be universal.
 
 The inclusion of feedback in SPH simulations is complicated by the so called overcooling
 problem(Katz~\citeyear{K92}). A number of methods has been devised to overcome this, based on the 
 return of energy as either thermal(Thacker \& Couchman~\citeyear{TC00}, Gerritsen~\citeyear{G97th}) 
 or kinetic( Navarro \& White~\citeyear{NW93}, Springel \& Hernquist~\citeyear{SH03}).
 None has been entirely satisfactory as the former requires the artificial suppression of
 cooling, while the latter suffers from too strong effects for feedback, unless an unrealistically 
 low SN efficiency is adopted or the affected particles are decoupled from the gasdynamics.
 Here, we employ a new method. It is based on the creation at the location of young stellar 
 clusters of a zero mass SPH particle. Its contribution to gas forces on neighbouring gas 
 particles is determined by taking the formal zero mass limit with constant particle energy 
 $E_{SN}$ of the usual SPH equations of motion(see Pelupessy et al.~\citeyear{PWIXX} for more details). 
 The amount of feedback( the energy $\rm E_{SN}$ given to the zero mass SPH particle) is an important 
 and uncertain parameter; we estimate it by multiplying the number of type II supernovae per mass of 
 stars formed, $\rm 0.009 \msun^{-1}$, with the effective energy per supernova, 
 $\rm \approx 10^{50} erg$, thus assuming that $90\%
 $ of the energy is radiated away, a value which comes from more detailed simulations of the 
 effect of supernova and stellar winds on the ISM (Silich et al.~\citeyear{SFPT96}), and is also 
 used in other simulations of galaxy evolution( e.g. Semelin \& Combes~\citeyear{SC02}, 
 Springel \& Hernquist~\citeyear{SH02}, Buonomo et al.~\citeyear{BCCL00}). This energy is then 
 gradually added to the supernova particle over a period of $\rm 30~Myr$, the lifetime of SN progenitors,
 after which it is ceases to exist. This model for the mechanical energy input from stellar clusters
 resembles the one adopted by Oey \& Clarke~(\citeyear{OC97}), who have shown that it is fully consistent
 with the differential size distribution of superbubbles in a.o. the SMC and Holmberg II. We do not 
 include the effect of SN Ia, which are mainly important for the chemical enrichment.
 
\subsection{ Initial conditions}

 Here we will restrict ourselves to follow the evolution of an isolated dwarf galaxy in an 
 attempt to understand the interplay between star formation and the ISM. These processes play a role for 
 dwarf galaxies in general, and indeed in all places were star formation is happening. It is important 
 to keep in mind, however, that dwarf galaxy evolution is easily influenced by external factors such
 as infall of gas, collisions, and recently receiving much attention, ram pressure stripping by the inter galactic
 medium and tidal stripping by galactic potentials( see e.g. Marcolini et al.~\citeyear{MBD03}, Mori
 \& Burkert~\citeyear{MB00} and Mayer et al.~\citeyear{MGC01}, Pasetto et al.~\citeyear{PCC03} for recent 
 work on these). Nevertheless the isolated case remains very relevant as it allows a straightforward 
 comparison with suitably selected samples of observed galaxies which exist in the literature(e.g. 
 van Zee~\citeyear{Z01}), unaffected by the uncertainties of additional external influences. The Understanding of
 the complex processes of star formation and feedback gained from the isolated case can then be aplied 
 to more general evolution scenarios.
 
 We thus setup a model dwarf galaxy resembling current dIrrs. Although these exhibit a wide range of 
 morphologies, they are very similar in their averaged radial profiles to scaled down versions of 
 ordinary disk galaxies. Hence we take for the initial condition a three component model for a 
 small disc galaxy, consisting of a gas disk, a stellar disk and a dark halo, modelled loosely 
 on the properties of DDO 47 and similar dIrrs. The gas disk we construct with a radial surface 
 density profile,
 \begin{equation}
 \rm 
 \Sigma= \Sigma_g/(1+R/R_{g}),
 \end{equation}
 with central density $\rm \Sigma_g=0.01~\msun/kpc^2$ and radial scale
 $\rm R_g=.75~kpc$, truncated at 6 kpc. It is somewhat involved, due to the thickness of the gas disk, to 
 solve its vertical structure exactly. Hence, before we start the simulation proper, we set up 
 the galaxy with an quadratically rising gas scaleheight and we it run isothermally($\rm T=10^4~K$) and 
 without star formation and feedback for some time until it settles in equilibrium. This results in a 
 scaleheight set by hydrostatic equilibrium of $\rm h_g=100~pc$ at the center to $\rm h_g=1~kpc$ at 
 $\rm R=6~kpc$.An exponential stellar disk, 
 \begin{equation}
 \rm 
 \rho_{disk}(R,z) = \frac{\Sigma_0}{2 h_z} \exp(-R/R_d) sech^2(z/h_z)
 \end{equation}
 with central surface density $\rm \sigma_0=0.3~\msun/kpc^2$, $\rm R_d=0.5~kpc$ and 
 vertical scale height $\rm h_z=0.2~kpc$, is constructed as in Kuijken \&  Dubinski~(\citeyear{KD95}). 
 The total mass of the gas is $\rm M_g=2 \times 10^8 \msun$ and the total mass of the stellar disk is
 $\rm M_d=1.5 \times 10^8 \msun$.  Both the gas disk and stellar disk are represented by  $\rm N=10^5$ 
 particles. The ages of the initial population of stars are distributed according to a constant SFR of
 0.007 $\rm \msun/yr$.
 
 Dwarf galaxies are amongst the most dark matter dominated objects, exhibiting dark to luminous matter 
 ratios of 10-100. Their rotation curves are best fit by flat central density cores (
 Flores \& Primack~\citeyear{FP94}, Burkert~\citeyear{B95}). Therefore we take a halo profile   
 \begin{equation}
 \label{eq_dh}
 \rm \rho_{halo}(r)= \rho_0 \frac{\exp(-r^2/r_c^2)}{1+r^2/\gamma^2}
 \end{equation}
 with core radius $\rm \gamma=2~kpc$, cutoff radius $\rm r_c=20~kpc$ and central density 
 $\rm \rho_0=2 \times 10^7~\msun/kpc^3$, for a total mass of $\rm M_{halo} = 15 \times 10^9 \msun$ and a 
 peak rotation velocity of about 50 km/s. The profile (\ref{eq_dh}) is very similar to the 
 Burkert~(\citeyear{B95}) profile for dwarf galaxies, differing in its asymptotic behavior 
 for $\rm r \rightarrow \infty$. This will only give significant deviations well outside 
 the region of interest for our simulations. We represent the dark halo by a static potential.
 This is deemed sufficient for the dynamical modelling presented here, as we evolve the galaxy in
 isolation and the perturbations in the gaseous and stellar disk are expected to have only minor
 impact on the halo structure. Furthermore, discreteness noise of a particle halo can induce bars
 to form, and excite spiral structure or bending modes in the stellar and gaseous disk( Kuijken
 \& Dubinski~\citeyear{KD95}, Hernquist~\citeyear{H93}). 

\section{ Simulation Results}

 \begin{figure*}
 \resizebox{!}{\vsize}{\includegraphics{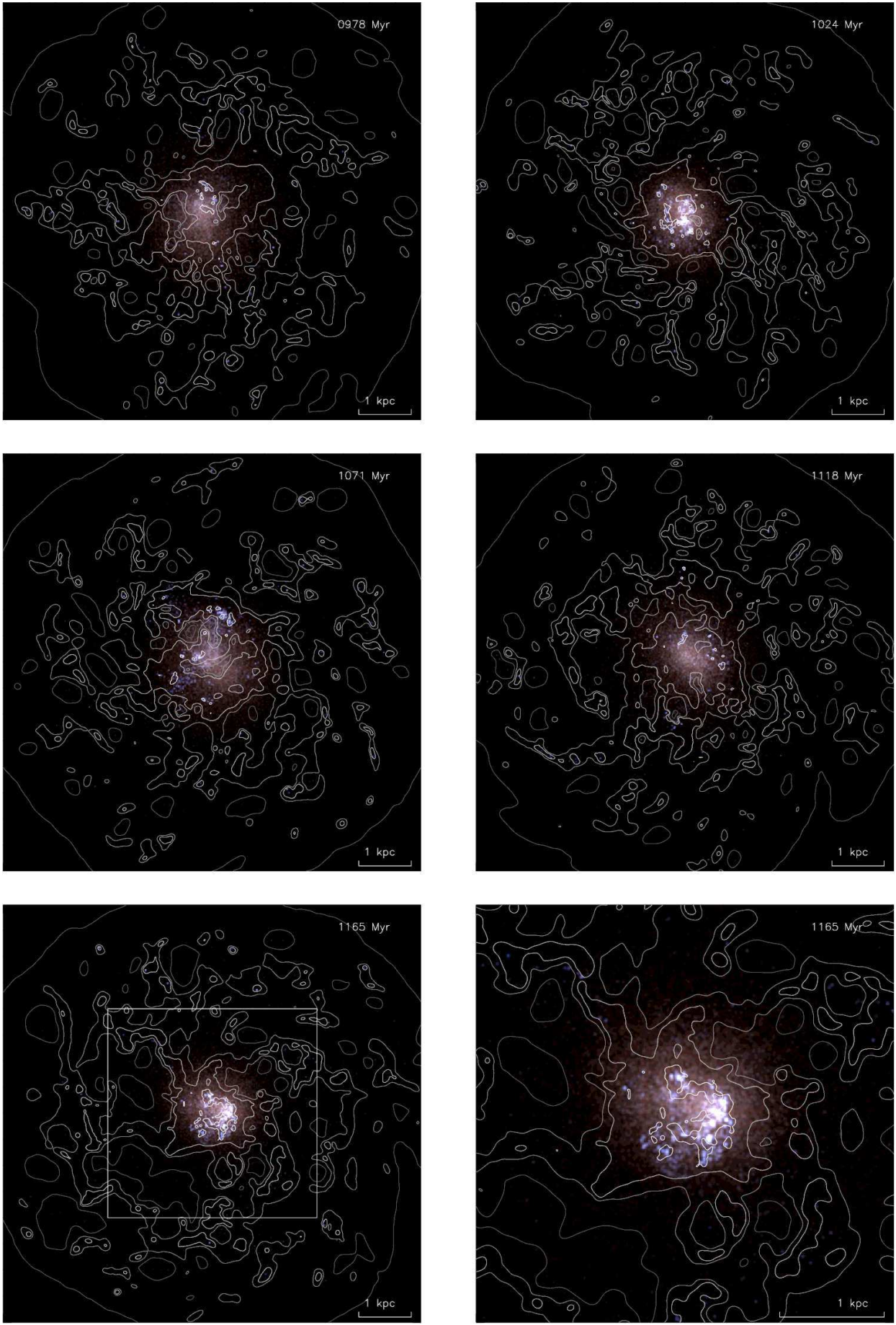}}
 \caption{ Simulated dwarf galaxy after $\approx 1$ Gyr. Shown are
 UBV composite pictures overplotted with HI density contours at different times. Last frame shows 
 the central regions at $\rm t= 1165~Myr$. HI contour 
 levels are at $\rm N_{HI}= 10^{20}$, $2.1~10^{20}$, $3.6~10^{20}$ and $2~10^{21}~{\rm cm^{-2}}$ }
 \label{UBVfig}
 \end{figure*}

 \begin{figure*}
 \resizebox{\hsize}{!}{\includegraphics{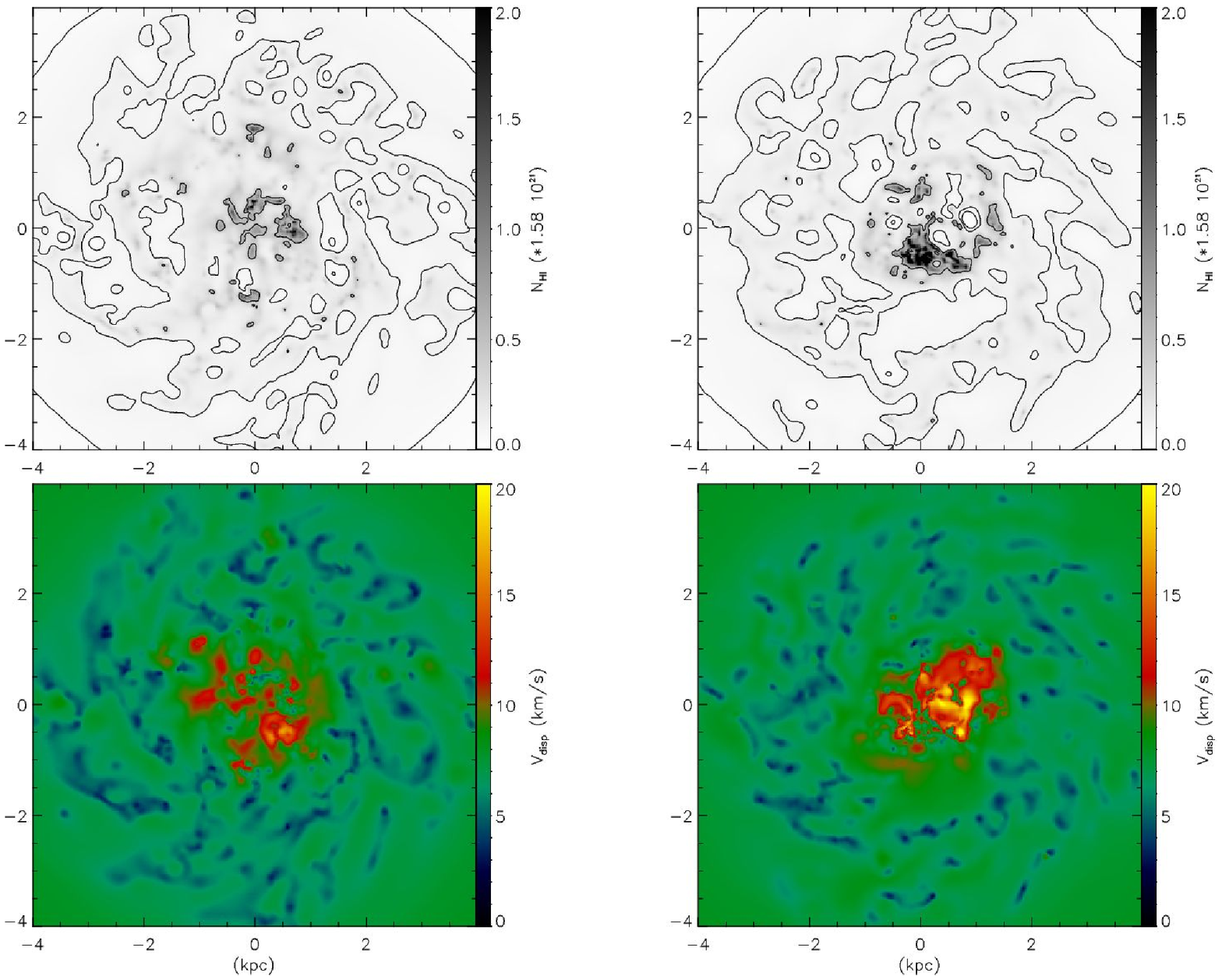}}
 \caption{ HI surface density \& velocity dispersion map in quiescent phase(left, taken
  at 1165 Myr) and in burst phase( right, at 1211 Myr).}
 \label{dispfig}
 \end{figure*}

 In Fig.~\ref{UBVfig} a range of snapshots of the simulated dwarf galaxy is shown, taken
 after about 1 Gyr of evolution and spaced about 50 million years apart. Plotted are UBV 
 composite images with HI contours overplotted. If we look at the contours and maps of HI 
 of the upper panels of Fig.~\ref{dispfig} we see that the HI distribution of the galaxy is 
 dominated by holes of varying sizes. Dense shells of HI combine to form big HI clouds, which 
 correspond to the sites of intense star formation. Sometimes structures resembling spiral arms
 form, but generally there are no spiral density waves in gas or stars, and, due to the low 
 shear, a flocculent spiral structure also does not develop. As the cold clouds 
 move about in the galaxy they are in turn destroyed by mechanical feedback from the stars 
 to perpetuate the cycle of star formation, cloud destruction and formation. Clearly the 
 ISM of the galaxy is in a very dynamical state. This can only be fully appreciated 
 while looking at the time sequence of the complete simulation. \emph{ The ISM of 
 dwarf irregulars is continously stirred and material from different radii is mixed 
 through the action of supernovae and stellar winds}. While this may have been expected
 beforehand from energy considerations, it is nonetheless an important and often 
 overlooked fact. As such it should not be surprising that irregulars often show 
 very little radial metallicity gradients (apart from the fact that metals may be lost from
 hot outflows). Note that the presence of the hydrodynamics of the ISM is a fundamental
 difference between classical SSPSF and phenomological models and our full dynamical 
 simulations.  
  
 If we look at the UBV images we see that the appearance of the galaxy varies with time. 
 Episodes of strong star formation( e.g. at 1024 Myr) are followed by quiescent phases( 
 1071 and 1118 Myr ). During star formation bursts the galaxy is dominated by a few 
 very active star formation sites, during quiet times the galaxy fades and seems relatively 
 featureless. If we compare the stellar and HI distribution of Fig.~\ref{UBVfig} to observed 
 systems, it is striking how similar these are. For example Holmberg II and NGC 4214, which have
 been studied in great detail (Puche et al.~\citeyear{PWBR92}, Walter \& Brinks~\citeyear{WB01}), 
 show an HI distribution with a similar number of holes of similar sizes. Also the extent of 
 structure in the gas outside the optical disk is reproduced in the simulations. Large regions of
 star formation are associated with the highest contours of HI (corresponding to a column density 
 of about $\rm N=2 \times 10^{21} cm^{-2}$), which is also found in comparisons of optical and HI morphology 
 for observed systems.

 As we have a complete representation of the neutral phases of the ISM we can also compare our
 simulation with the HI kinematics of observed systems. In Fig.~\ref{dispfig} we have 
 plotted maps of the HI density and the (line of sight) velocity dispersion of the gas. 
 The gas typically has random velocities of the order of 3-10 km/s, with regions of higher 
 (up to 20 km/s) velocity dispersion associated with expanding bubbles, in good agreement with
 observations (Stil~\citeyear{S99}, Young et al.~\citeyear{YL97}).
  
\begin{figure*}
 \resizebox{\hsize}{!}{\includegraphics{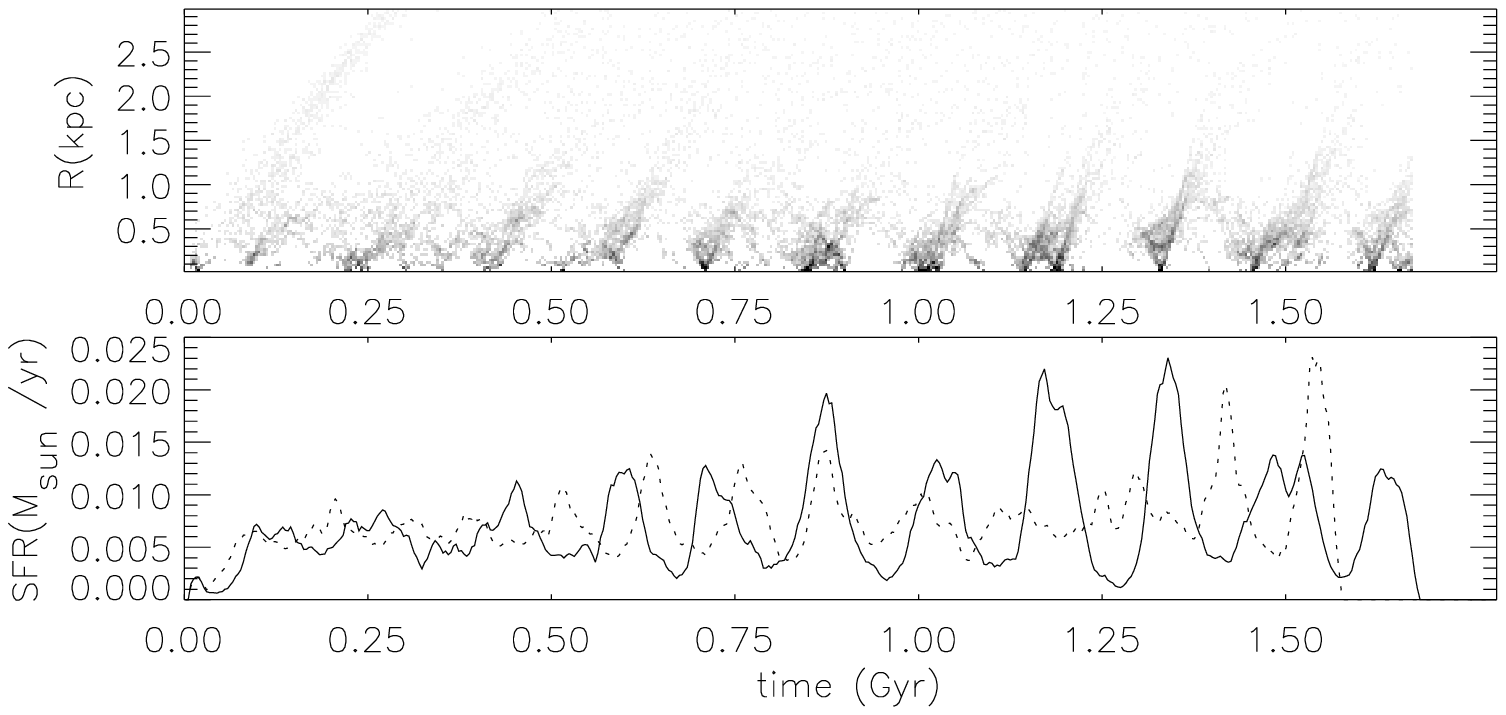}}
 \caption{ Star Formation History of a simulated dwarf galaxy. Upper panel: density plot of 
 the azimutally averaged star formation rate. Lower Panel: total star formation rate. The 
 dotted line indicates the SFR for a run with $50\%$ reduced feedback strength.}
 \label{SFfig}
 \end{figure*}

\begin{figure}
 \resizebox{\hsize}{!}{\includegraphics{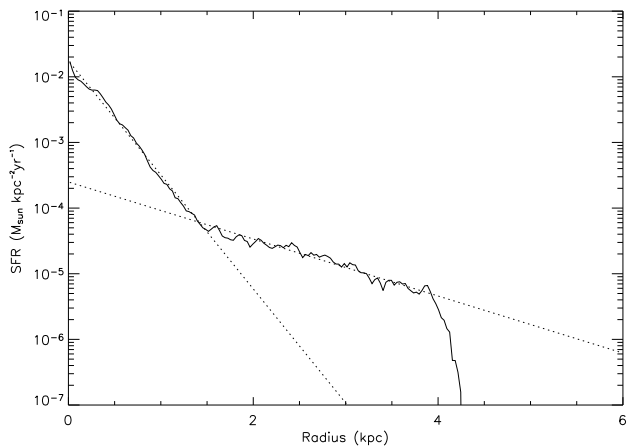}}
 \caption{ Radial profile of the mean star formation density( drawn line, dotted lines are 
 exponentials with scalelengths of .25 and 1. kpc).} 
 \label{SFDfig}
 \end{figure}

 The evolution of the SFR is shown in Fig.~\ref{SFfig}. It shows a gray scale plot of
 the temporal evolution of the (azimuthally averaged) star formation density and below
 it the resulting total SFR. We see that after an initial transient period the galaxy 
 settles in a mode of periodically varying star formation, with a minimum SFR of about
 $\rm 0.003 \msun/yr$ while the peak SFR is about a factor of 10 higher. Also plotted are the results
 for a run with reduced feedback strength. In this case variations are smaller in amplitude,
 and quasi-periodic. Even lower feedback strengths would probably quench the variations
 completely, but in that case morphology and kinematics of the gas would no longer match 
 observed systems. In the upper panel of Fig.~\ref{SFfig} we see the pattern of propagating star formation.
 The azimuthally averaged density plot may give the false impression that star formation is happening in 
 inward and outward moving rings; however visual inspection shows that it moves around in patches and partial 
 rings: if some region starts to form stars, nearby dense regions will be triggered to form stars, often moving 
 the star formation in a particular direction along a dense filament or bridge. Furthermore we see that periodic 
 bursts happen only relatively close to the centre of the galaxy, whereas in the outer parts 
 stars form at a more or less constant low rate. This behaviour is expected (see Ehlerov{\' a} \& 
 Palou{\v s}~\citeyear{EP02}) because star formation can only be triggered in the expanding shells around 
 the holes induced by feedback if the column density is high enough, $\rm N \approx 10^{20 - 21}~cm^{-2}$. So 
 we see that the galaxy has three different SF regimes: an outer region where the gas density is generally 
 not high enough for the CNM to form (beyond the so called thermal cut off, see Elmegreen \& 
 Parravano~\citeyear{EP94} and Gerritsen \& Icke \citeyear{GI97}) so SF will be strongly 
 supressed there; a region at intermediate radii where the gas density is high enough to be thermally
 unstable, and where star formation proceeds at a steady pace; and a central region that is both 
 thermally unstable and unstable to shell instabilities, where the dominant mode of SF is triggered 
 star formation. Fig.~\ref{SFDfig} shows a plot of the resulting radial dependence of the star forming density,
 clearly visible are the three different regions with different star forming behaviour. The bursting region is
 confined within 1.5 kpc, while star formation extends further out, stopping abrubtly at $\rm \approx 4 kpc$, 
 well short of the edge of the gasdisk. Interestingly, De Blok \& Walter~(\citeyear{BW03}) found evidence for 
 similar low level star formation outside the optical disk of NGC 6822.  

 The cyclical pattern of star formation raises a number of further questions: What is 
 the driver of these variations?  What determines the period and amplitude of the oscillation? 
 Why is star formation apparently synchronized over the whole galaxy? While some previous 
 models of galaxies have shown variations of the star formation rate, these were either 
 connected to an imposed time scale (the duration of supernova feedback, 
 Wada \& Norman~\citeyear{WN01}) or due to random scatter of a probabilistic model (Gerola 
 et al.~\citeyear{GSS80}). Significant quasi random variations in SF rate were found by 
 Carraro et al.~(\citeyear{CCGL01}) to happen during formation of dwarf ellipticals (dEs), in 
 their case caused by local depletion of the starforming gas reservoir, happening on 
 timescales of $\approx 10^{8-9}$ yr. Similar variations driven mainly by internal proccesses were 
 found by Pasetto et al.~(\citeyear{PCC03}) even for dIrrs subjected to tidal stirring. 
 In our case the timescale of the variations is related to a physical time scale: the period of 
 the oscillation is about 165 Myr, which is slightly larger than double the dynamical time scale 
 of the darkhalo, ($\rm t_{dyn}=\sqrt{\frac{3 \pi}{16 G \rho}}=80~Myr$). This suggests that supernova 
 feedback provides the kick for the oscillation, expelling gas and increasing the velocity dispersion 
 and thus inflating the gas disk, after which the gas falls back for the next cycle on a 
 dynamical time scale. The period is thus determined by the density of the halo (the main 
 mass component of dwarf galaxies), while the amplitude is related to the strength of the 
 supernova feedback. Note that this would not be sufficient to account for the regularity of 
 the variations: in general one would expect that gas from different parts of the gas disk 
 would experience kicks of different strength, and would fall back on different time scales, 
 giving rise to a more chaotic pattern of star formation. However dwarf galaxies 
 are in solid body rotation and thus their potential corresponds to a harmonic oscillator, which 
 means that material flung out of the disc falls back in the same time, irrespective of 
 location and speed, synchronizing the SFR over the whole galaxy. This effect certainly adds 
 to the phenomenon shown in Fig.~\ref{SFfig}, namely that the overall star formation cycle looks 
 remarkably regular.

\section{Star formation: comparison with observations }

 When we compare the star formation in our model and observed SFR (for example in 
 Van Zee~\citeyear{Z01}) we see that the agreement is quite good: our values fall well 
 within the range of observed rates of $0.001 \msun/yr$ to $0.1 \msun/yr$, although 
 admittedly the star formation rate in our model does depend fairly strongly on poorly 
 constrained quantities such as the shape of the IMF, the amount of energy released in 
 SN and stellar winds, and the various parameters regulating star formation. It is 
 however encouraging that we can use reasonable values for the parameters and get SFR 
 that are close to observed rates. 

\begin{figure}
 \resizebox{\hsize}{!}{\includegraphics{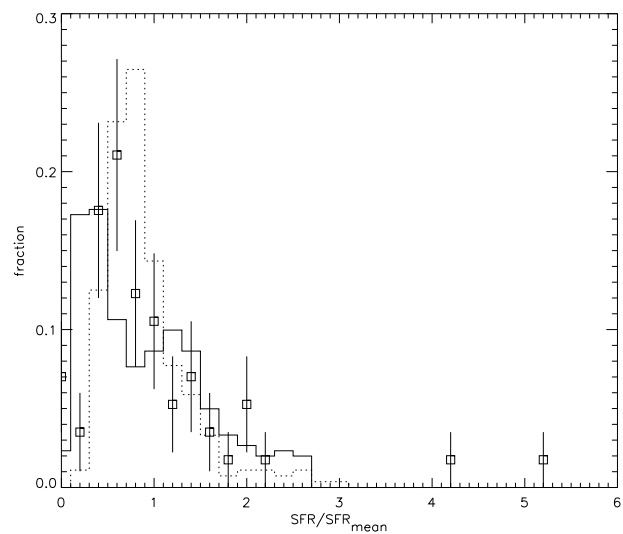}}
 \caption{ distribution of $\rm SFR/\langle SFR \rangle_t$ for a sample of isolated dwarfs
 (squares with error bars), a simulation with full feedback(drawn line) and a simulation with $50\%$
 feedback( dotted line). Data is taken from Van Zee~\citeyear{Z01}.}
 \label{b_fig}
 \end{figure}

 In our model the star formation rate is seen to vary over time. This is difficult to verify in
 real dwarf galaxies. One would have to derive accurate SFH over hundreds of Myr for galaxies, and 
 variations of within a factor $\approx 3$ of the mean as observed in our
 model would not be too conspicous in the data. For most dwarf galaxies the long term star 
 formation histories determined from global colours indicate an approximately constant SFR. 
 Detailed examinations of nearby resolved dwarfs do show evidence of variations 
 in the SFR (Dohm-Palmer et al.~\citeyear{DSG98},~\citeyear{DSM02}) of the correct order of magnitude. 
 Strong additional evidence for variations in the SFR is the observed scatter in the ratio of current 
 star formation to past star formation: the Scalo $b$ parameter. If the variations in our model were generic 
 for isolated dwarf galaxies, this would explain the observed distribution of $b$. To see this we 
 have plotted in Fig.~\ref{b_fig} the histograms of $\rm SFR/\langle SFR \rangle_t$ for our simulation 
 together with the distribution of $b$ for a sample of isolated dwarf galaxies as derived by 
 Van Zee~(\citeyear{Z01}). Clearly these two are qualitatively the same. There are two main differences:
 (1) the distribution of our model extends to higher and lower ratios and (2) the galaxy 
 sample contains galaxies with $b>4$, which are extreme bursting systems. The first difference 
 would be remedied by decreasing the strength of feedback. We also have plotted the results 
 for a run with $50\%
 $ less feedback. In this case the amplitude of variations is somewhat too small(although both
 are in good agreement with the observed distribution given the uncertainties inherent in deriving it,
 see discussion in Van Zee~\citeyear{Z01}). For the second point we should mention 
 that our simulation really only samples a limited number of cycles; the bursts may be rarer
 events that have not (yet) happened in the simulation. Ofcourse it also possible that high powered
 bursts are due to outside influences, such as interactions or infalling gas clouds, ingredients of 
 galaxy evolution not present in our model.
 
 The distribution of $b$ is a highly degenerate indicator for the SF history, and we are aware 
 that the above comparison presupposes a number of properties for the population of dwarf 
 galaxies: the assumption that the dwarf galaxies form a homogeneous population; variations
 in SFR are assumed to be of similar strength and shape, 
 independent of galaxy properties( although the $b$ distribution is independent of the period); 
 and that the variations in SFR are periodic. Nevertheless we think that this is the simplest 
 explanation for the observed scatter and that our model captures a number of essential features
 of the variations in star formation given the characteristic shape of the observed $b$ 
 distribution, namely: moderate bursting, long periods of low SF, and the 'peakiness' of the 
 SF enhancements. It is also difficult to see what would induce non-periodic variations in 
 isolated galaxies, especially since no properties independent from the SFR that correlate 
 with the SFR (Van Zee~\citeyear{Z01}) have been found. 

\section{ Discussion \& Conclusions}

 The model we employed to calculate the evolution of a single dwarf galaxy is quite extensive and 
 complicated and necessarily contains a number of free parameters, some of which we tried to constrain 
 using observational data or theory( as in the case of the supernova energy), yet some 
 remain only constrained by physical intuition. Therefore, it is worthwhile to 
 summarize again the features of observed dwarf systems successfully reproduced by the model: 1) the
 morphology and kinematics of the HI distribution, 2) the two phase structure of the ISM, 2) the spatial 
 pattern of star formation, 3) the star formation rate, 4) consistency with the observed distribution of
 Scalo $b$ parameter. These provide strong and independent checks of our modelling approach as they 
 result from the intricate interplay of the model for the ISM we employed, the scheme for star formation 
 and the method of returning mechanical energy to the ISM. These successes give some confidence that the 
 model captures the essentials of dwarf galaxy evolution and to its predictive power. 

 Comparing our model with recent simulations of dwarf galaxy evolution( e.g. Berzcik et al.~\citeyear{BHTS03}, 
 Mayer et al.~\citeyear{MGC01}, Andersen \& Burkert~\citeyear{AB00}, Pasetto et al.~\citeyear{PCC03}, 
 Mori et al.~\citeyear{MYTN97},~\citeyear{MYN99}, Carraro et al.~\citeyear{CCGL01}) we see that our work differs
 from previous work in two crucial aspects: 1) the modelling of the physics of the neutral ISM, 2) the treatment 
 of supernova feedback. We follow the evolution of the WNM and CNM of the ISM explicitly, this process being the upper 
 part of a cascade leading down to star formation. Star formation in our model happens in cold( $\rm T \la 200~K$) 
 and dense ($\rm n=1-10~cm^{-3}$) gas. These are realistic sites for star formation as we know that molecular clouds
 are embedded in neutral envelopes. The further stages of star formation, molecular cloud formation and collapse, are
 not included as they require prohibitive resolution and the inclusion of additional physics. These processes are only implicit
 in our star formation recipe, but at least our methods follows collapsing gas to the point that it has experienced 
 a transition to a cold, dense state from which it may be trusted to form stars with rates and efficiencies that are  
 constrained by observations. Some authors have tried to bypass this problem by formulating a phenomenological ISM 
 model in terms of 'sticky' particles representing molecular clouds (Andersen \& Burkert~\citeyear{AB00}), 
 sometimes in addition to a smooth SPH component representing the WNM (Semelin \& Combes~\citeyear{SC02},
 Berczik et al.~\citeyear{BHTS03}). They also succesfully reproduce a selfregulated ISM, including some
 effects, like evaporation of molecular clouds that are probably not well represented in our model. 
 In these models stars form from the cloud particles, imposing a Schmidt law, not, as in our model, from 
 the consideration of the the instabilities in the ISM. Furthermore, our model has a consistent representation of 
 the ISM linking phases by physical processes, rather than prescriptions. 
 
 Simulations of dwarf galaxy evolution that include supernova feedback have typically included this as a local heating
 term. This method for implementing supernova feedback is not effective in forming the structures associated with 
 stellar feedback and simulations using it do not show any effect of SN feedback. This is a well understood numerical
 artefact and some authors have devised methods to prevent the radiative loss of mechanical energy that is the root of 
 the problem (Springel \& Hernquist~\citeyear{SH03}, Thacker \& Couchman~\citeyear{TC00}, Gerritsen \& Icke~\citeyear{GI97}, 
 the problem is also not exclusive to SPH type codes see e.g. Fragile et al.~\citeyear{FMAL03}). They have not applied their methods to 
 the evolution of dIrrs so a direct comparison is not yet possible. However, the notion that \emph{the energy input 
 from stellar winds and SN of young stellar clusters is crucial for the understanding the structure and
 kinematics of the ISM of dIrrs} is borne out by our simulations: if feedback is not included, the ISM stays 
 in a smooth disk with very low random motions( 2-3 km/s). The self propagating mode of
 star formation is absent in that case. The dwarf galaxy will only poorly resemble a real dIrr.  
 
 Chemical enrichment is not yet included in our code. We can estimate the importance of a 
 changing chemical composition for our simulation. The total amount of metals $\rm M_z$ produced during the 
 simulation is $\rm M_Z \approx 0.015 \Delta M_{\star} = 1.6 \times 10^5 \msun$. This will raise 
 the metallicity  of the galaxy, under the assumption that the metals will be well mixed, with at most $25\%
 $, which is not entirely insignificant, although to first approximation the thermal evolution is 
 independent of metallicity as both the cooling and UV heating scale with Z. Note also that
 metals may be lost from the galaxy in galactic winds. It does mean that we should include chemical 
 enrichment to follow the evolution for longer timescales and especially if we want to investigate 
 extremely metal poor galaxies. In principle the inclusion of chemical enrichment will also add further 
 constrains comparing with observations, very important to asses the long term evolution of the model.
 Both these points, however, do not alter the conclusion for simulations as presented here.
  
 The methods we use can be applied more general. Questions that we plan to address are 
 for example the following:
 \begin{itemize} 
 \item{
 In view of the recent realization that there is a deficit of observed small galaxies as 
 compared to predictions of $\Lambda$CDM models of galaxy formation
 (Klypin et al.~\citeyear{KKVP99}) it is interesting to consider what happens for galaxies of
 progressively smaller mass. We are in the process of running a grid of models exploring
 this question, but some effects may be clear from the preceding: for smaller galaxies
 variations will be on longer time scales (scaling as $\rho_{halo}^{-.5})$ and bigger in
 amplitude (because the feedback becomes relatively stronger as the escape velocity 
 decreases). Ultimately halos will be too small to retain their ISM, leaving them bare.
 The mass ranges for which this happens, the influence of other galactic parameters and the
 timescales on which these processes take place will be of interest to validate cosmological 
 and galaxy formation models. 
 }
 \item{
 We also plan to investigate a possible relation between dwarf irregulars and dwarf ellipticals. 
 Although recent simulations(Mayer et al.~\citeyear{MGC01}, Pasetto et al.~\citeyear{PCC03}) have 
 shown for satellite galaxies that a transition from dIrr to dE or dwarf Spheroidal (dSph) is 
 plausible as a result of the action of tidal fields, it has not been conclusively determined whether 
 a transition from dIrr to dE is to be expected in general. Various people have put forward 
 arguments in favour (Davies \& Phillipps~\citeyear{DP88}) and against (Bothun et al.~\citeyear{BMCM86}, 
 Marlowe et al.~\citeyear{MMH99}) such a descendancy for the dE. Although the results 
 presented here do not immediately illuminate this question, we think that additional 
 simulations of the sort presented here, testing a wider range of galactic properties and 
 following the evolution for longer time scales may answer whether this is a viable
 scenario and whether we can put the various classes of small galaxies into an unified 
 evolutionary framework. 
 }
 \end{itemize}

 In summary, our model suggests that it is possible that a large part of the current dIrr
 population is in a quasi-periodic burst mode of star formation. The scatter in observed
 properties in this picture is mainly due to the fact that galaxies are observed at different phases
 of their evolution. The main difference between our model and classical SSPSF models is that
 in our models variations are due to the interplay of stellar feedback and gasdynamics, the galaxy
 being periodically stirred by bursts of star formation after which a quiescent period occurs during
 which gas falls back to the disk. In classical SSPSF models the variations are due to the
 fact that star formation is described by a correlated stochastic process and the small size of the
 galaxies, which induces large statistical variations. Our model predicts that the amplitude 
 of the variations depends on the strength of the feedback and that the period depends on the 
 dynamical time scale. 

 \begin{acknowledgements}
The authors want to thank Jeroen Gerritsen and Roelof Bottema for providing
software and analysis tools. FIP wants to thank Padelis Papadopoulos for contributing to
this paper by lending his time to many discussions on dwarf galaxies.   
\\
 This work was sponsored by the stichting Nationale Computerfaciliteiten
(National Computing Facilities Foundation) for the use of supercomputer 
facilities, with financial support from the Nederlandse Organisatie voor
Wetenschappelijk Onderzoek (Netherlands Organisation for Scientific Research,
NWO). FIP was supported by a grant from NWO.
 \end{acknowledgements}

\end{document}